\documentclass[aps,prd,superscriptaddress,12pt,showpacs,notitlepage]{revtex4}
\usepackage{amsmath,amssymb,mathrsfs}
\usepackage{graphicx}

\newcommand{\MeV}{\textrm{MeV}}

\newcommand{\be}{\begin{equation}}
\newcommand{\ee}{\end{equation}}
\newcommand{\bea}{\begin{eqnarray}}
\newcommand{\eea}{\end{eqnarray}}
\newcommand{\pa}{\partial}

\newcommand{\Tr}{\textrm{Tr}}
\newcommand{\im}{{\rm i}}

\begin{document}
\title{Nuclear binding energies from a BPS Skyrme model}
\author{C. Adam}
\affiliation{Departamento de F\'isica de Part\'iculas, Universidad de Santiago de Compostela and Instituto Galego de F\'isica de Altas Enerxias (IGFAE) E-15782 Santiago de Compostela, Spain}
\author{C. Naya}
\affiliation{Departamento de F\'isica de Part\'iculas, Universidad de Santiago de Compostela and Instituto Galego de F\'isica de Altas Enerxias (IGFAE) E-15782 Santiago de Compostela, Spain}
\author{J. Sanchez-Guillen}
\affiliation{Departamento de F\'isica de Part\'iculas, Universidad de Santiago de Compostela and Instituto Galego de F\'isica de Altas Enerxias (IGFAE) E-15782 Santiago de Compostela, Spain}
\author{A. Wereszczynski}
\affiliation{Institute of Physics,  Jagiellonian University,
Reymonta 4, Krak\'{o}w, Poland}


\begin{abstract}
Recently, within the space of generalized Skyrme models, a BPS submodel was identified which reproduces some bulk properties of nuclear matter already on a classical level and, as such, constitutes a promising field theory candidate for the detailed and reliable description of nuclei and hadrons. Here we extend and further develop these investigations by applying the model to the calculation of nuclear binding energies.
Concretely, we calculate these binding energies by including the classical soliton energies, the excitation energies from the collective coordinate quantization of spin and isospin, the electrostatic Coulomb energies and a small explicit isospin symmetry breaking, which accounts for the mass difference between proton and neutron. The integrability properties of the BPS Skyrme model allow, in fact, for an analytical calculation of all contributions, which may then be compared with the semi-empirical mass formula. We find that for heavier nuclei, where the model is expected to be more accurate on theoretical grounds,  
 the resulting binding energies are already in excellent agreement with their physical values. This result provides further strong evidence for the viability of the BPS Skyrme model as a distinguished starting point and lowest order approximation for the detailed quantitative investigation of  nuclear and hadron physics.
\end{abstract}

\maketitle

\section{ Introduction} 

The Skyrme model \cite{skyrme}, \cite{manton} was introduced by Skyrme as a field theoretic realization of his concept of a "mesonic fluid" model for nuclei, motivated by the substantially homogeneous nature of nuclear matter. The nucleons were supposed to result from a kind of local twist or "vorticity" in this mesonic fluid, that is, to be described by topological solitons in a more modern language. This original idea received further support and general acceptance once it was found, about two decades later, that an effective theory of mesons may be derived from QCD in the limit of a large number of colors \cite{thooft}. So, the  Skyrme model is a nonlinear field theory which is supposed to describe the low-energy limit of strong interaction physics in terms of hadrons. The primary fields of the Skyrme model are mesons, whereas baryons appear as collective excitations, that is, topological solitons (skyrmions). These solitons have the characteristic feature that they can be classified by an integer-valued topological degree or winding number.
One important insight of Skyrme has been the proposal that this topological degree should be identified with the baryon number $B$, explaining in this way its conservation. 

For the specific case of two flavors, the field of the Skyrme model takes values in the group SU(2) (isospin). And as anticipated by Skyrme, one natural area of applications of the resulting theory is nuclear physics, providing in this manner a possible basis for a unified field theoretic description of nuclei and their properties. Indeed, shortly after the description of the nucleons (proton and neutron) in terms of the simplest skyrmion (the hedgehog solution) \cite{nappi}, the Skyrme model was used to describe the deuteron \cite{braaten} and some additional light nuclei \cite{carson}. More recently, nuclear excitation spectra have been studied within the Skyrme model, e.g., in \cite{wood}, with reasonable success. One common problem in the application of the standard version of the Skyrme model to nuclei is that the resulting nuclear binding energies are too large. Concretely, although there exists a lower energy bound linear in the baryon number (Faddeev-Bogomolnyi bound, see, e.g., \cite{rybakov-book}, \cite{manton-book}), skyrmion solutions do not saturate this bound. The energy (mass) of the simplest $B=1$ skyrmion (hedgehog) is about 23\% above the bound, whereas for higher $B$ this deviation is lowered, to less than 4\% in the limit of very large $B$ (see, e.g., \cite{manton-book}). As a consequence, the binding energies per baryon number of higher skyrmions (i.e., the energies needed to desintegrate higher skyrmions into their $B=1$ constituents) are quite high, on the level of 10\%, which is in striking contrast to the low binding energies of physical nuclei. A way of dealing with this problem in the standard Skyrme model is by a different renormalization of the coupling constants of the model for different nuclei.    

Recently, some of us found \cite{BPS-Sk} that there exists a certain Skyrme submodel ("BPS Skyrme model") which not only allows for a Bogomolnyi-type energy bound, but also contains infinitely many BPS soliton solutions (that is, soliton solutions which solve a first-order (BPS) equation) which saturate the bound. As already suggested in \cite{BPS-Sk} (see also \cite{marleau1}),
this observation leads to the rather natural proposal to treat the solitons of the BPS submodel as the leading order contributions to nuclear masses. This proposal receives further support from the observation that the submodel has the symmetries of an incompressible ideal liquid \cite{BPS-Sk}, \cite{fosco} and, therefore, serves as a field theoretic realisation of the liquid drop model of nuclei. Relatively small corrections to the nuclear masses (and, therefore, small but nonzero binding energies) may be produced by further small contributions, which may be incorporated  within the model in a completely natural fashion, that is, they are integral parts of the Skyrme theory itself. One should, e.g., extend the theory by
including additional terms into the classical energy functional. Further, in any case, one has to go beyond the classical solitons (e.g., via the usual collective coordinate quantization, and by including the electrostatic Coulomb energy, etc.) for an accurate description of nuclei. It is the purpose of the present article to advance in this second direction, and to determine the resulting nuclear binding energies. Concretely, we shall include the effects of the collective coordinate quantization of spin and isospin, the Coulomb energy and an explicit small breaking of the isospin symmetry. We will find that the resulting nuclear binding energies are already in very good agreement with the experimental values for heavy nuclei, demonstrating that the BPS Skyrme model together with standard Skyrme technology provide an excellent starting point for a detailed quantitative investigation of nuclear and low-energy strong interaction physics. An additional virtue of the BPS Skyrme model is that, due to its BPS nature, almost all calculations can be done analytically, which should be contrasted with the challenging numerical calculations required to find the solitons of the standard Skyrme model.  
We remark that an alternative proposal for a BPS Skyrme model has been developed in \cite{sutcliffe}.  
It is based on the inclusion of an infinite
tower of vector mesons which induces a flow to a conformal theory
(see also \cite{rho}).

\section{BPS Skyrme model}

The lagrangian of the standard Skyrme model consists of two terms, a term quadratic in first derivatives (the sigma model term ${\cal L}_2$) providing the kinetic energy of the pions, and the Skyrme term ${\cal L}_4$ which is quartic in first derivatives. The Skyrme term is needed to balance the scaling instability (avoid the Derrick theorem) such that soliton solutions may exist. 
The Skyrme model is meant to be a low-energy effective field theory for hadrons and nuclei, and, as such, in principle allows (and requires) the addition of many more terms, essentially all possible terms compatible with the basic symmetries of strong interactions. This is precisely what is done in a perturbative approach (chiral perturbation theory \cite{ChPT}). For a nonperturbative framework like the Skyrme model, however, some selection principle must be found which highlights the relevant physical effects and degrees of freedom and, further,  reduces the number of possible terms, rendering calculations feasible.  Concretely, we shall require that the lagrangian is no more than quadratic in time derivatives such that a standard Hamiltonian exists, which clearly is a desireable property for a theory supporting solitons. Together with the obvious requirement of Poincare invariance, this severely restricts the possible terms in the lagrangian. Essentially, only a potential term ${\cal L}_0$ and a certain term ${\cal L}_6$ sextic in derivatives may be added to the two standard terms ${\cal L}_2$ and ${\cal L}_4$. 
It should be noted that both of these terms have already been considered in generalizations of the original Skyrme model, where their inclusion was based on physical (phenomenological) arguments. The potential is obviously related to an explicit breaking of chiral symmetry and to the pion masses, whereas the sextic term may be related to vector meson exchange \cite{sextic}. 

So we are lead to the lagrangian
(our Minkowski metric conventions are ${\rm diag} (g_{\mu\nu}) = (+,-,-,-)$)
\be
{\cal L}={\cal L}_0 + {\cal L}_2 + {\cal L}_4 + {\cal L}_6
\ee 
where 
\be
{\cal L}_2 = \lambda_2 {\rm tr}\; \pa_\mu U \pa^\mu U^\dagger \equiv - \lambda_2 {\rm tr}\; L_\mu L^\mu \; ,
\quad {\cal L}_4 = \lambda_4 {\rm tr}\; \left( [L_\mu ,L_\nu ] \right)^2
\ee
and 
\be
{\cal L}_0 = -\lambda_0 V( U) \; , \quad {\cal L}_6 = -\lambda_6 \left( \epsilon^{\mu\nu\rho\sigma} {\rm tr}\; L_\nu L_\rho L_\sigma \right)^2 
\equiv -(24 \pi^2)^2 \lambda_6 {\cal B}_\mu {\cal B}^\mu .
\ee
Here $U: {\mathbb R}^3 \times {\mathbb R} \to {\rm SU(2)}$ is the Skyrme field and $L_\mu = U^\dagger \pa_\mu U$ is the left-invariant Maurer-Cartan current. Further, the $\lambda_n$ are non-negative coupling constants, and ${\cal B}_\mu$ is the topological or baryon number current giving rise to the topological degree (baryon number) $B\in {\mathbb Z}$,
\be
{\cal B}^\mu = \frac{1}{24 \pi^2 } \epsilon^{\mu\nu\rho\sigma} {\rm tr}\; L_\nu L_\rho L_\sigma \; ,\quad B= \int d^3 x {\cal B}^0 .
\ee
 The terms ${\cal L}_2$, ${\cal L}_4$ and ${\cal L}_6$ are invariant under the chiral transformations $U \to WUW'$, $W,W' \in $ SU(2). 
The potential term, on the other hand, breaks this chiral symmetry. 
We shall assume from now on that the potential only depends on ${\rm tr}\; U$, i.e., $V(U)= v({\rm tr}\; U)$, then it is still invariant under the diagonal (isospin) subgroup $U\to WUW^\dagger$. Further, we assume that the potential is non-negative, $V(U)\ge 0$, and has one unique vacuum at $U={\bf 1}$, i.e., $V(U={\bf 1})=0$. 

Our BPS Skyrme model is the limit $\lambda_2 = \lambda_4 =0$ of the above theory, with lagrangian (where, for convenience, we introduce the new coupling constants $\lambda_6 = \lambda^2 /(24)^2$ and $\lambda_0 = \mu^2$) 
\be
{\cal L}_{06} = - \pi^4 \lambda^2  {\cal B}_\mu {\cal B}^\mu - \mu^2 V(U) .
\ee
Clearly, this simple model by itself cannot provide a detailed description of all strong interaction physics. Due to the absence of the sigma model term ${\cal L}_2$ there is, e.g., no perturbative pion dynamics. 
The model is based on two terms which are related to collective, nonperturbative properties of strong interactions (chiral symmetry breaking for the potential, and Skyrme field topology for the sextic term), so it might be expected to be relevant whenever nonperturbative properties should be important like, for instance, in regions of not too small baryon density (as is the case, e.g., inside nuclei). Further, the model may   
provide a good starting point (lowest order approximation) in these cases, where a more detailed investigation then will require the inclusion of further effects. This proposition is bolstered by the crucial observation that the energy functional of the theory for static configurations,
\be \label{BPS-en-funct}
E = \int d^3 x \left( \pi^4 \lambda^2  {\cal B}_0^2 + \mu^2 V(U) \right)
\ee
 has both a Bogomolnyi bound and infinitely many BPS solutions saturating this bound \cite{BPS-Sk}, \cite{fosco}, \cite{ferr}. To derive the bound, it is useful to recall that the target space SU(2) as a manifold is just the three-sphere ${\mathbb S}^3$ and, further, the topological charge density three-form ${\cal B}_0 d^3 x$ is (up to a constant) the pullback (under the map $U: {\mathbb R}^3 \to {\mathbb S}^3$) of the volume form $d\Omega$ on ${\mathbb S}^3$, i.e., ${\cal B}_0 d^3 x = (1/(2\pi^2 ))U_*(d\Omega )$. Then it is not difficult to find the Bogomolnyi bound \cite{BPS-Sk}, \cite{fosco}, \cite{speight}
\be
E \ge  2\pi^2 \lambda \mu |B| <\sqrt{V}>_{{\mathbb S}^3} \; , \quad <\sqrt{V}>_{{\mathbb S}^3} \equiv \frac{1}{2\pi^2} \int_{{\mathbb S}^3} d\Omega \sqrt{V}
\ee
(where $<\sqrt{V}>_{{\mathbb S}^3}$ is the average value of $\sqrt{V}$ on the target space ${\mathbb S}^3$) and the BPS equation
\be
\pi^2 \lambda {\cal B}_0 \pm \mu \sqrt{V} =0.
\ee
Another important property of the energy functional (\ref{BPS-en-funct}) is its infinite-dimensional group of symmetries \cite{BPS-Sk}, \cite{fosco}, among which there are the volume-preserving diffeomorphisms (VPDs) on base space (i.e., the symmetries of an incompressible ideal liquid). 

At this point, we have to make several choices. First of all, for simplicity we choose the standard Skyrme potential
\be \label{Sk-pot}
V = \frac{1}{2} {\rm tr} \; ({\bf 1}-U) = 1-\cos \xi ,
\ee
where
\be
U= \cos \xi +\im \sin \xi \; \vec n \cdot \vec \tau \; , \quad \vec n^2 =1,
\ee
and $\vec \tau$ are the Pauli matrices (in this article we use the roman letter i for the imaginary unit, in order to avoid confusion with the isospin quantum number $i$). We remark that
without the term ${\cal L}_2$ there is no direct relation between the standard Skyrme potential and the pion mass, so other choices are possible \cite{potentials}. Secondly, we have to choose the shapes (symmetries) of our skyrmion solutions. Here we shall choose an axially symmetric ansatz, but due to the many symmetries of the energy functional (\ref{BPS-en-funct}), there exist solutions with rather arbitrary shapes.  Indeed, introducing spherical polar coordinates $(r,\theta, \phi)$ in base space, and $\vec n = (\sin \chi \cos \Phi ,\sin \chi \sin \Phi , \cos \chi )$, the BPS equation may be written like \cite{fosco}
\begin{equation}
-\frac{\lambda}{2\mu} \frac{\sin^2 \xi}{\sqrt{V}} \sin \chi d\xi d\chi d\Phi = \mp r^2 \sin \theta dr d\theta d\phi .
\end{equation}
The obvious ansatz
\begin{equation} \label{ansatz}
 \chi = \theta, \;\;\; \Phi = n\phi , \quad \xi = \xi (r) 
\end{equation}
then leads to $B=n$ and to the first order ODE for $\xi (r)$
\begin{equation}
 \frac{n\lambda}{2\mu}  \frac{\sin^2 \xi}{\sqrt{V}} \frac{d\xi}{dr}  =  - r^2 .
\end{equation}
For the standard Skyrme potential (\ref{Sk-pot}), the solution obeying the right boundary conditions $\xi (0)=\pi$, $\xi (\infty)=0$ is \cite{BPS-Sk}
\begin{equation}
\xi = \left\{
\begin{array}{lc}
2 \arccos \frac{r}{ R_n} & \; \; \; \; r \in \left[0,  R_n \right] \\
0 & \hspace*{-0.3cm} r \geq  R_n 
\end{array} \right. ,
\quad R_n \equiv \left( \frac{2\sqrt{2} \lambda |n|}{\mu} \right)^\frac{1}{3}
\end{equation}
that is, the skyrmion for baryon number $n$ is a compacton with radius $R_n$. We remark that the compacton radius grows like the third root of the baryon number, which exactly reproduces the experimental behaviour of the radii of physical nuclei. For later use we also want to display the following expression, valid inside the compacton radius,
\be
\sin^2 \xi \; \xi_r = -8\frac{r^2}{R_n^3} \sqrt{1-\left( \frac{r}{R_n}\right)^2 } .
\ee 

Next, we have to decide which effects (contributions) to include in our binding energy calculations. We shall, at present, not include (small) contributions from the terms ${\cal L}_2$ and ${\cal L}_4$. The reason is as follows. As they are small by assumption, the main contributions of these terms would just slightly increase the classical soliton energies. For large baryon number, where our model is most reliable, the full soliton energies and, therefore, also these contributions must grow linearly in the baryon charge (they cannot grow faster, because the "soliton" then would be unstable, and they cannot grow slower because of the Bogomolnyi bounds). But they can then be taken into account effectively by a slight renormalisation of the coupling constants of the restricted (BPS) model. As we have to fit the coupling constants to experimental values in any case, the leading effect of these terms is therefore immaterial. We remark that in \cite{marleau}, where similar calculations are performed, the terms ${\cal L}_2$ and ${\cal L}_4$ are, nevertheless, taken into account perturbatively. We shall comment on this issue in the conclusions.

\section{Nuclear binding energies}

For the static energy (mass) $E$ of a nucleus, we shall take into account the following contributions,
\begin{equation} \label{StaticEnergy}
E=E_0 + E_{rot} + E_C + E_I.
\end{equation}
where $E_0$ is the classical soliton energy, which for the standard Skyrme potential is \cite{BPS-Sk}
\begin{equation}
E_0 = \frac{64 \sqrt 2 \pi}{15} \mu \lambda |n|.
\end{equation}
 $E_{rot}$ comes from the collective coordinate quantization of rotations and iso-rotations, giving rise to spin and isospin of the nuclei. $E_C$ is the electrostatic Coulomb energy of the nucleus, whereas $E_I$ is a contribution from a small, explicit isospin breaking, taking into account the mass difference between proton and neutron.

\subsection{Spin and isospin contributions}

The (semi-classical) quantization of rotations and iso-rotations is obviously required for a consistent description of nuclei, because both spin (angular momentum) and the third component of the isospin, $i_3 = (1/2)(Z-N)$, are relevant quantum numbers of nuclei (here $Z$ is the number of protons, and $N $ the number of neutrons).  As usual, the rigid rotor quantization of spin and isospin proceeds by introducing the rotational and iso-rotational degrees of freedom about a static soliton via
\be \label{coll-coord}
U(t,\vec x)= {\rm A}(t)U_0(R_{\rm B}(t) \vec x){\rm A}^\dagger (t)
\ee
(${\rm A, B} \in $ SU(2), $R_{\rm B} \in$ SO(3), $U_0$ $\ldots$ soliton), where the variables $a^n(t)$ and $b^n(t)$ parametrizing $A$ and $B$ are treated as time-dependent mechanical coordinates. This expression is then inserted into the lagrangian $L = \int d^3 x {\cal L}$, where only time derivatives provide additional terms (after all, rotations and iso-rotations are symmetries). In a next step, one transforms from the generalized velocities $\dot a^n$, $\dot b^n$ to the canonical momenta, and from the mechanical lagrangian $L(a^n ,b^n,\dot a^n ,\dot b^n)$ to the Hamiltonian. In a last step, the coordinates and canonical momenta are then interpreted as quantum mechanical variables and momenta fulfilling the corresponding commutation relations. The result is the standard quantized rigid rotor both for spin and for isospin, and different nuclei are identified with different eigenstates of the rigid rotor, i.e., 
\be \label{nuc-state}
|X\rangle = |jj_3 l_3\rangle |ii_3k_3\rangle
\ee
where $X$ is a nucleus, $\vec J $ ($\vec L$) is the space-fixed (body-fixed) angular momentum,
$\vec I$ ($\vec K$) is the space-fixed (body-fixed) isospin angular momentum, and $j, j_3, l,l_3$ and $i ,i_3 ,k, k_3$ are the corresponding eigenvalues. Finally, $|jj_3l_3\rangle$ and $|ii_3k_3\rangle$ are the eigenstates of the rigid rotor (Wigner D functions) for spin and isospion, respectively.

A slight complication in this rigid rotor quantization is related to the symmetries of the skyrmion $U_0$. Indeed, for soliton solutions $U_0$ with some symmetries, certain combinations of the transformations $A$ and $B$ will act trivially on $U_0$. If these combinations correspond to (one or several) continuous one-parameter families, then the corresponding combinations of collective coordinates do not show up at all in the quantum mechanical Hamiltonian $\widehat H_{rot}$ and, further, these transformations should act trivially on the nuclei described by the soliton $U_0$, (that is, the infinitesimal generators acting on $|X\rangle $ should give zero). If these symmetries of $U_0$ are just discrete transformations, then they do not reduce the number of collective coordinates, but they still imply some nontrivial constraints on the nuclear wave functions $\vert X \rangle$, the so-called Finkelstein-Rubinstein constraints \cite{Fin-Rub} (in general, these imply that only certain linear combinations of the product states (\ref{nuc-state}) are possible, although for the axially symmetric ansatz considered here the allowed wave functions may always be written as product states). A detailed discussion of the Finkelstein-Rubinstein constraints for the standard Skyrme model may be found in \cite{krusch}.      

For the ansatz (\ref{ansatz}) used here, the corresponding moments of inertia are well-known \cite{nappi}, \cite{braaten}, \cite{hou-mag}.
Concretely, for $n=B=1$, the resulting skyrmion (hedgehog) is spherically symmetric, i.e., an arbitrary rotation can be undone by an iso-rotation (and vice versa). As a consequence, only three of the six collective coordinates (either spin or isospin) appear, where we choose spin for concreteness. Further, the body-fixed moments of inertia tensor is proportional to the identity,
\begin{equation} \label{Jota-sym}
{\cal J}_{ij} =\delta_{ij} {\cal J}, \; \; {\cal J} = \frac{4\pi}{3} \lambda^2 \int dr \sin^4 \xi \, \xi_r^2 = \frac{2^8 \sqrt{2} \pi}{15\cdot 7} \lambda \mu \left( \frac{\lambda}{\mu}\right)^\frac{2}{3}
\end{equation} 
and the resulting quantum mechanical hamiltonian is the one of a spherical top, ${\cal H}_{rot} = \left( 1/(2{\cal J}) \right) \vec L^2 = \left( 1/(2{\cal J}) \right) \vec J^2$ leading to the energy
\begin{equation}
E _{\rm rot}= \frac{1}{2{\cal J}} \hbar^2 j(j+1) .
\end{equation}

For the axially symmetric ansatz (\ref{ansatz}) for $n=B> 1$, the quantum mechanical hamiltonian essentially consists of
two copies (spin and isospin) of a symmetric top (rigid rotor with ${\cal J}_{ij} = {\cal J}_i \delta_{ij}$, and generically ${\cal J}_1 = {\cal J}_2 \not= {\cal J}_3$) with
\begin{equation}
{\cal H}_{\rm sym-top} = \frac{L_1^2 + L_2^2 }{2{\cal J}_1} + \frac{L_3^2 }{2{\cal J}_3} = \frac{\vec J^2 }{2{\cal J}_1 } + \left( \frac{1}{2{\cal J}_3} - \frac{1}{2{\cal J}_1} \right) L_3^2 .
\end{equation}
The axial symmetry implies that a rotation about the three-axis (by an angle $\varphi$) can be undone by an isospin rotation (by an angle $n\varphi$), so the corresponding generator ($L_3$ or $K_3$) should be taken into account only once (concretely we choose $K_3$). The resulting energy is
\begin{equation} \label{E-rot}
E_{\rm rot} = \frac{\hbar^2}{2} \Bigl( \frac{j (j+1)}{{\cal J}_{1}} + \frac{i (i+1)}{{\cal I}_{1}} 
+ \bigl( \frac{1}{{\cal I}_{3}} - \frac{1}{{\cal I}_{1}} - \frac{n^2}{{\cal J}_{1}} \bigr) k_3^2 \Bigr) 
\end{equation}
where ${\cal I}_{ij} = {\cal I}_i \delta_{ij}, {\cal I}_1 = {\cal I}_2 \not= {\cal I}_3$ is the isospin moments of inertia tensor, and
\bea
&& {\cal I}_3 = \frac{4\pi}{3} \lambda^2 \int dr \sin^4 \xi \, \xi_r^2 = |n|^{-\frac{1}{3}}{\cal J} \\
&&  {\cal I}_1 = \frac{3n^2 + 1}{4} {\cal I}_3 \; , \quad  {\cal J}_1 = {\cal J}_3 = n^2 {\cal I}_3 
\eea
and ${\cal J}$ is defined in (\ref{Jota-sym}).
We were able to guess the energy expression (\ref{E-rot}) with the help of the standard rigid rotor quantization and the axial symmetry of the skyrmion, but for later use it is still useful to sketch the explicit calculation. 
Indeed, inserting (\ref{coll-coord}) into the Lagrangian, the resulting quantum mechanical Lagrangian (which is equal to the Hamiltonian ${\cal H}_{\rm rot}$, because we ignore the constant soliton mass) may be expressed as a quadratic form in the spin ($\omega_k$) and isospin ($\Omega_k$) angular velocities
\be \label{H-rot-omega}
{\cal H}_{\rm rot} = \frac{1}{2} \Omega_j {\cal I}_{jk} \Omega_k -\Omega_j {\cal K}_{jk} \omega_k + \frac{1}{2} \omega_j {\cal J}_{jk} \omega_k 
\ee
where 
\begin{equation}
{\rm A}^\dagger \dot{\rm A} = \im \, \vec \Omega \cdot \frac{\vec \tau}{2}.
\end{equation}
and
\begin{equation}
(\dot{R}_{\rm B})_{ik} R_{\rm B}^{-1}{}_{kj} = - \epsilon_{ijk} \omega_k.
\end{equation}
Further, ${\cal K}_{jk} = {\cal K}_j \delta_{jk}$, ${\cal K}_1 = {\cal K}_2 =0$, ${\cal K}_3 = n {\cal I}_3$ is the "mixed" moments of inertia tensor.
Now, the transformation  to body-fixed spin and isospin angular momenta proceeds as usual,
\be \label{vec-K}
\vec K = \frac{\partial {\cal H}_{\rm rot}}{\partial \vec \Omega} = \left( {\cal I}_1 \Omega_1 , {\cal I}_1 \Omega_2 , {\cal I}_3 (\Omega_3 -n\omega_3 ) \right)
\ee
\be \label{vec-L}
\vec L = \frac{\partial {\cal H}_{\rm rot}}{\partial \vec \omega} = \left( {\cal J}_1 \omega_1 , {\cal J}_1 \omega_2 , -n {\cal I}_3 (\Omega_3 -n\omega_3 ) \right) .
\ee
Resolving these expressions for $\vec \omega$ and $\vec \Omega$, inserting them into (\ref{H-rot-omega}) and replacing the angular momentum operators by their eigenvalues, we precisely recover (\ref{E-rot}). We remark that here we use the angular velocity and body-fixed angular momentum sign conventions of \cite{braaten}. This implies that the body-fixed angular momenta obey the normal commutation relations $[K_1 ,K_2] = \im K_3$ etc. On the other hand, the body-fixed angular momenta are then related to the space-fixed ones by {\em minus} the corresponding rotation. 

Finally, the axial symmetry implies
\be
(L_3 + n K_3) |X\rangle =0 \quad \Rightarrow \quad  l_3 + n k_3 =0.
\ee
This constraint, together with the obvious inequality $j\ge |l_3|$, leads to unacceptably large angular momenta for physical nuclei for $k_3 \not= 0$. We, therefore, assume $k_3 =0$ in what follows. But this assumption implies that the axially symmetric ansatz cannot be used for nuclei with odd baryon number $B$, because such nuclei are fermions with half-odd integer values of $k_3$. We shall, therefore, restrict our discussion for the axially symmetric solitons to nuclei with even $B=n$. We remark that we differ in this respect from the discussion in \cite{marleau}.  
Further, from now on we assume $i=|i_3|$, for the following reason.
We want to compare our calculated excitation energies with the binding energies of the most abundant nuclei with the same quantum numbers $B$, $j$ and $i_3$. But these nuclei typically correspond to the most tightly bound ones, so their excitation energies should take on the minimum possible values, i.e., obey the condition $i=|i_3|$. Assuming this, we get for (even) $B=n>1$ 
\be
E_{rot} = \frac{105}{512 \sqrt 2 \pi} \frac{\hbar^2}{\lambda^2 \big( \frac{\mu}{\lambda n} \big)^{1/3}} \Bigl( \frac{ j (j+1)}{n^2} +\frac{4|i_3|(|i_3|+1)}{3n^2 +1} \Bigr) .
\ee
For $B=n=1$ (where $j=\frac{1}{2}$), the well-known result is
\begin{equation}
E_{\rm rot} =  \frac{105}{512 \sqrt 2 \pi} \frac{3}{4} \frac{\hbar^2}{\lambda^2 \left( \frac{\mu}{\lambda} \right)^{1/3}} .
\end{equation}

\subsection{Coulomb energies}

The Coulomb energy contribution is
\begin{equation}
 E_C=\frac{1}{2 \varepsilon_0} \int d^3 x d^3 x' \frac{\rho(\vec r) \rho(\vec r\,')}{4 \pi|\vec r - \vec r\,'|}
\end{equation}
where $\rho$ is the electric charge density of the nucleus. This charge density is the expectation value w.r.t. nuclear wave functions $| X \rangle$ of the corresponding electric charge density operator \cite{cal-wit} (which, in the underlying QCD description, incorporates not only the minimal coupling, but also the effects of the chiral anomaly)
\be
\hat{\rho} =\frac{1}{2} {\cal B}^0 + {\mathbb J}_3^0 
\ee
where ${\cal B}^0$ is the topological charge density and $ {\mathbb J}_3^0 $ is the time component of the third isospin current density operator ${\mathbb J}^\mu_a$. Again, $\hat \rho$ is calculated by inserting the spin- and isospin-rotated soliton (\ref{coll-coord}) into the above expression and by interpreting the collective spin and isospin coordinates as quantum mechanical variables. The first term, $(1/2){\cal B}^0$, is, in fact, proportional to the identity operator, 
\begin{equation} \label{B0}
{\cal B}^0 = - \frac{n}{2 \pi^2 r^2} \sin^2 \xi \, \xi_r ,
\end{equation}
because the topological charge density is invariant under spin and isospin rotations (contains no time derivatives).  For the second term a complication arises due to the fact that, with the term ${\cal L}_6$ present, 
${\mathbb J}^0_3 = {\mathbb J}^0_3 (\vec \Omega, \vec \omega, a^n)$ depends not only on the spin and isospin angular velocities, but also explicitly on the corresponding isospin collective coordinates, so a Weyl ordering is required (see \cite{Ding} for details; we remark that here we differ from \cite{marleau}). Indeed,  
\begin{equation}
{\mathbb J}^\mu_3=-\frac{\im \lambda^2 \pi^2}{4} \epsilon^{\mu \nu \alpha \beta} {\cal B}_\nu \Tr \left[ \frac{\tau_3}{2} (\partial_\alpha U U^\dagger \partial_\beta U U^\dagger - \partial_\alpha U^\dagger U \partial_\beta U^\dagger U) \right],
\end{equation}
so
\begin{equation}
{\mathbb J}^0_3=-\frac{\im \lambda^2 \pi^2}{4} \epsilon^{0imn} {\cal B}_i \Tr \left[\frac{\tau_3}{2} (\partial_m U U^\dagger \partial_n U U^\dagger - \partial_m U^\dagger U \partial_n U^\dagger U) \right] .
\end{equation}
Here, the first factor ${\cal B}_i$ contains a time derivative and will therefore depend on $\vec \Omega$ and $\vec \omega$, whereas the second term, $\Tr ( \tau_3 \ldots )$ is non-invariant under isospin transformations and depends, therefore, on the corresponding collective coordinates. A long but straight-forward calculation leads to the explicitly Weyl-ordered expression
\begin{equation} \label{iso-dens-op}
{\mathbb J}_3^0 = - \frac{\lambda^2}{4r^2} \xi_r^2 \sin^4 \xi \bigl( (R_i \Omega_j + \Omega_j R_i)  \,  {\cal A}_{ij}  + (R_i \omega_j + \omega_j R_i)  \, {\cal B}_{ij} \bigr) ,
\end{equation} 
where $\vec R$ is a unit vector constructed from the isospin collective coordinates $a_0 , \vec a$,
\bea
&& {\rm A}^\dagger \tau_3 {\rm A} = R_i \tau_i \; , \quad {\rm A} = a_0 + i \vec a \vec \tau \; , \quad a_0^2 + \vec a^2 =1 \quad \Rightarrow \\
&& R_1 = 2(a_0 a_2 + a_1 a_3), \quad R_2 = 2(a_2 a_3 - a_0 a_1), \quad R_3 = a_0^2 - a_1^2 -a_2^2 + a_3^2.
\eea
For later use we remark that, at the same time, $R_i$ is the third component of the rotation matrix from the body-fixed to the space-fixed coordinates (w.r.t. the isospin rotation), i.e., $R_i \equiv (R_{\rm A})_{3i}$, where
\be
(R_{\rm A})_{jk} = \frac{1}{2} \Tr \left( \tau_j {\rm A}^\dagger \tau_k {\rm A}\right) ,
\ee 
which implies, e.g., the useful relation 
\be \label{I3-rel}
I_3 = -\sum_i R_i K_i 
\ee
(with the minus sign due to our conventions)
between body-fixed and space-fixed isospin angular momenta. Further, ${\cal A}_{ij}$ and ${\cal B}_{ij}$ are matrices which depend on the angles $\theta $ and $\phi$ of the spherical polar coordinates. As the explicit expressions are rather lengthy, we relegate them to appendix A. 

Now we have to evaluate the nuclear matrix elements $\langle X | {\mathbb J}^0_3 | X\rangle$, where several simplifications occur. We replace the angular velocities by the angular momentum operators using (\ref{vec-K}), (\ref{vec-L}) and use that the matrix elements $\langle X | R_i K_j | X \rangle$ etc. are zero for $i\not= j$, see appendix B. For $i=j$, on the other hand, the operators commute and the Weyl-ordering may be ignored. Further, the matrix elements $\langle X | L_1 | X\rangle $ and $\langle X | L_2 | X\rangle $ are zero (the presence of the operators $R_i$ does not change this, because $\vec L$ and $\vec R$ act in different spaces). Using ${\cal B}_{33} = - n {\cal A}_{33}$, we therefore get for the matrix element
 \begin{equation}
\langle X |{\mathbb J}_3^0 | X \rangle = - \frac{\lambda^2}{2r^2} \xi_r^2 \sin^4 \xi \frac{1}{{\cal I}_3} \langle X | \left( \frac{4}{3n^2 +1}(R_1 K_1 {\cal A}_{11} + R_2 K_2   {\cal A}_{22})  + R_3 K_3 {\cal A}_{33} \right) | X \rangle .
\end{equation} 
In a next step, we use that as a consequence of the symmetries of the nuclear wave functions $|X\rangle$ it holds that $\langle X| R_1 K_1 |X\rangle =
\langle X| R_2 K_2 |X\rangle$ so that we may replace both of them by $(1/2) (\langle X| R_1 K_1 |X\rangle + \langle X| R_2 K_2 |X\rangle $. 
Adding and subtracting, in addition, $R_3 K_3$ to complete for $I_3$ (see Eq. (\ref{I3-rel})), we arrive at
\bea
\langle X |{\mathbb J}_3^0 | X \rangle &=&  \frac{\lambda^2}{2r^2} \xi_r^2 \sin^4 \xi \frac{1}{{\cal I}_3} \langle X | \left( \frac{2(n^2 + \cos^2 \theta )}{3n^2 +1}I_3 - \frac{(n^2 +1)(1-3\cos^2 \theta )}{3n^2 +1} R_3 K_3 \right) | X \rangle \nonumber \\
&=& \frac{\lambda^2}{2r^2} \xi_r^2 \sin^4 \xi \frac{1}{{\cal I}_3} \left( \frac{2(n^2 + \cos^2 \theta )}{3n^2 +1}i_3 - \frac{(n^2 +1)(1-3\cos^2 \theta )}{3n^2 +1} \langle X|R_3 K_3 |X\rangle \right) 
\eea
It may be checked easily that the contribution to the electric charge is just $i_3$, i.e., $\int d^3 x \langle X| {\mathbb J}_3^0 |X\rangle =i_3$. The matrix element $\langle X|R_3 K_3 |X\rangle $ does not contribute even for nuclei with $k_3 \not= 0$, because its prefactor integrates to zero.  
For $n=B>1$, we may ignore this term even for the electric charge densities, because the corresponding nuclear wave functions we consider obey $K_3 |X\rangle =0$.
So, the electric charge density for $n>1$ is
\bea
\rho &=& \frac{1}{2} {\cal B}^0 + \langle X| {\mathbb J}_3^0 |X\rangle \nonumber \\
&=& -\frac{n}{4\pi^2 r^2 }\sin^2 \xi \; \xi_r + \frac{\lambda^2 i_3}{r^2 {\cal I}_3} \sin^4 \xi \; \xi_r^2 \frac{n^2 + \cos^2 \theta}{3n^2 +1} \\
&=& \frac{2n}{\pi^2 R_n^3 } \sqrt{ 1-\frac{r^2 }{R_n^2 } } + \frac{105 \, i_3}{8\pi R_n^3 }\frac{n^2 + \cos^2 \theta }{3n^2 +1}
\frac{r^2}{R_n^2} \left(1-\frac{r^2}{R_n^2} \right) .
\eea
For the calculation of the Coulomb energy one now has to perform the usual multipole expansion of the Coulomb potential \cite{marleau},
\be 
\frac{1}{4\pi | \vec r - \vec r' |} = \sum_{l=0}^\infty \sum_{m=-l}^l \frac{1}{2l+1}\frac{r_<^l}{r_>^{l+1}}Y^*_{lm}(\theta' ,\phi')Y_{lm} (\theta ,\phi)
\ee 
($r_< = {\rm min}(r,r')$, $r_> = {\rm max}(r,r')$) 
and expand the charge density into spherical harmonics,
\begin{equation}
 \rho (\vec r) = \sum_{l,m} \rho_{lm}(r) Y^{*}_{lm}(\theta,\phi),
\end{equation}
then the Coulomb energy can be expressed as
\begin{equation}
 E_C= \sum^\infty_{l=0} \sum^l_{m=-l} U_{lm}
\end{equation} 
where
\begin{equation} \label{Ulm}
 U_{lm}=\frac{1}{2 \varepsilon_0} \int^\infty_0 dr r^{-2l-2}|Q_{lm}(r)|^2
\end{equation}
and
\begin{equation}
 Q_{lm} (r) = \int^r_0 dr' r'^{l+2} \rho_{lm}(r')
\end{equation}
as may be checked easily. In our case, only two spherical harmonics contribute, 
\bea
&& \rho (\vec r) = \rho_{00}(r) Y_{00} + \rho_{20} (r) Y_{20} \nonumber \\
&& \rho_{00} (r)= \frac{4n}{\pi^\frac{3}{2}R_n^3} \sqrt{1-\frac{r^2}{R_n^2}} + \frac{35 \, i_3}{ 4\sqrt{\pi} R_n^3} \frac{r^2}{R_n^2} \left( 1-\frac{r^2}{R_n^2} \right) \\
&& \rho_{20} (r) = \frac{7\sqrt{5}}{2\sqrt{\pi}R_n^3}\frac{i_3}{3n^2 +1}\frac{r^2}{R_n^2} \left( 1-\frac{r^2}{R_n^2} \right) .
\eea
So, finally, the Coulomb energy is $E_{\rm C}=U_{00} + U_{20}$ and, after performing the integrations, the explicit expression reads, for $n=B>1$
\begin{eqnarray}
&E_{\rm C}=& \frac{1}{\sqrt{2} \pi\varepsilon_0} \Bigg( \frac{\mu}{\lambda n} \Bigg)^{1/3}
\Bigg( 
\frac{128}{315 \pi^2 }  n^2 + \frac{245}{1536 }  n \; i_3 + \nonumber \\
&& + \frac{805}{5148 }  i_3^2 
 + \frac{7}{429 }  \frac{i_3^2}{(1+3n^2)^2} \Bigg) .
\end{eqnarray}
For the hedgehog skyrmion solution $n=1$, the electric charge density has already been calculated in \cite{BPS-Sk},
\be
\rho (\vec r) = \frac{2}{\pi^2 R_1^3}\sqrt{1-\frac{r^2}{R_1^2}} \pm \frac{35}{16 \pi R_1^3 }\frac{r^2}{R_1^2} \left( 1-\frac{r^2}{R_1^2} \right) 
\ee
(the plus sign is for the proton, and the minus sign for the neutron). The resulting Coulomb energies are
\begin{eqnarray}
\mbox{proton:} \quad &E_C^{\rm p} =& 
\frac{1}{\sqrt{2} \pi\varepsilon_0} \Bigg( \frac{\mu}{\lambda } \Bigg)^{1/3} \Bigg(
\frac{128}{315 \pi^2}  + \frac{156625}{1317888} \Bigg) 
 \\
\mbox{neutron:} \quad &E_C^{\rm n} =&  
\frac{1}{\sqrt{2} \pi\varepsilon_0} \Bigg( \frac{\mu}{\lambda } \Bigg)^{1/3} \Bigg(
\frac{128}{315 \pi^2}  - \frac{53585}{1317888} \Bigg) .
\end{eqnarray} 

\subsection{Isospin breaking}
The Coulomb energy results of the last section would imply that the proton is heavier than the neutron, in striking contrast to reality. This contradiction is resolved by the fact that isospin is not an exact symmetry of strong interactions. On a microscopic level this follows from the mass difference between up and down quark, but within the Skyrme model it should result from isospin-breaking terms in the effective pion Lagrangian, 
like, e.g., mass terms which lead to slightly different masses for the charged and uncharged pions. The collective coordinate quantization then has to be done for this new Skyrme lagrangian with the isospin-breaking terms included. A detailed discussion of this issue is beyond the scope of the present article and will be given elsewhere (for related discussions see, e.g., \cite{iso-breaking}). Here we just take into account the leading order effect of the isospin breaking, which is obvious for physical reasons. Indeed, in leading order the isospin breaking part of the quantum hamiltonian of the collective coordinate quantization should just give a slightly higher mass to each neutron (a slightly smaller mass to each proton) and still commute with $I_3$ (which remains a good quantum number). The hamiltonian obviously is ${\cal H}_{\rm I} = a_{\rm I} I_3$ with the resulting energy contribution
\begin{equation}
E_{\rm I} = a_{\rm I} i_3 \quad {\rm where} \quad a_{\rm I}<0 .
\end{equation} 
In principle, it should be possible to calculate the constant $a_{\rm I}$ from the microscopic theory, but here we shall treat it as a free parameter. 

\section{Explicit binding energy calculations}
The idea now is to calculate explicit numerical values for the masses of nuclei from our model and to compare with the known experimental values.  
For this purpose, first of all we have to determine numerical values for the three parameters $\lambda $, $\mu$ and $a_{\rm I}$. Concretely, we fit to the nuclear masses of the proton, neutron, and the nucleus with magical numbers $^{138}_{\hspace{0.14cm}56}{\rm Ba}$, with masses
\begin{eqnarray}
M_{\rm p} &=& 938.272 \; \MeV  \\
M_{\rm n} - M_{\rm p} &=& 1.29333 \; \MeV  \\
M(^{138}_{\hspace{0.14cm}56}{\rm Ba}) & =& 137.894 \; \textrm u \quad \textrm{where} \quad \textrm{u} = 931.494 \; \MeV .
\end{eqnarray}
Further, we need the numerical values of some universal constants,
\begin{eqnarray}
\hbar &=& 197.327 \; \textrm{MeV  fm}  \\
\varepsilon_0 &=& \frac{1}{e} 8.8542 \cdot 10^{-21} \frac{1}{\textrm{MeV  fm}} \\
e &=& 1.60218 \cdot 10^{-19} .
\end{eqnarray} 
The fit then leads to the parameter values
\begin{equation}
\lambda \mu =   48.9902 \; \MeV , \;  \left( \frac{\mu}{\lambda} \right)^\frac{1}{3} =  
0.604327 \; \textrm{fm}^{-1}, \;
a_{\rm I} = -1.68593 \; \MeV .
\end{equation}
With these values, we now may determine the masses (energies $E_X$) of many more nuclei where
\begin{equation}
E_X = E_{\rm sol} + E_{\rm rot} + E_{\rm C} + E_{\rm I} .
\end{equation}
The main contribution to nuclear masses, however, is well-known to stem from the masses of the constituent protons and neutrons, so it is more instructive to determine (and plot), instead, the nuclear binding energies
\begin{equation}
E_{{\rm B},X} = Z E_{\rm p} + N E_{\rm n} - E_X ,
\end{equation} 
i.e., the differences between the masses of the constituents and the actual nuclei. We remark that the isospin-breaking term $a_{\rm I}i_3$ does not contribute directly to the binding energies (it cancels out). It contributes, of course, indirectly, because the fitted values of the parameters $\lambda $ and $\mu$ depend on it. 

Now we slightly change notation to
\be i_3 = \frac{1}{2}(Z-N) \; , \quad n=B\equiv A = Z+N
\ee
where $A$ is the atomic weight number of nuclear physics, then the binding energy for a given proton number $Z$, atomic weight number $A$ and spin $j$ may be expressed like
\begin{eqnarray} 
&E_{{\rm B},X}(A,Z,j) =& a_1 A + a_2 Z - a_3 A^{5/3} - a_4 A^{2/3} Z - a_5 A^{-1/3} Z^2 \nonumber \\
&& -a_6 \frac{A^{1/3}}{1+3 A^2} \, (A - 2 Z) - a_7\frac{A^{1/3}}{1+3 A^2} \, (A - 2 Z)^2  \nonumber \\
&& - a_8 \frac{A^{-1/3}}{(1+3 A^2)^2} \, (A - 2 Z)^2 - a_9 A^{-5/3} j (j+1)  \label{bind-en}
\end{eqnarray}
where
\begin{eqnarray}
&& a_1 =  10.0503 \; \MeV, \; 
a_2 =  0.400307 \; \MeV, \;
a_3 =  1.26027 \cdot 10^{-3} \; \MeV, \nonumber \\
&& a_4 =  0.100077 \; \MeV, \;
a_5= 0.384881 \; \MeV, \;
a_6 =  26.7974 \; \MeV, \nonumber \\
&& a_7 = 13.3987 \; \MeV, \;
a_8 = 0.0100404 \; \MeV, \;
a_9 =  13.3987 \; \MeV  .
\end{eqnarray}
Here, $a_1$ receives contributions from the classical soliton energy of the nucleus and from the masses of protons and neutrons, $a_2$ is nonzero because of the different Coulomb energies of proton and neutron, whereas $a_3$ - $a_5$ and $a_8$ stem from the Coulomb energy of the nucleus.
Further, $a_6$ and $a_7$ come from the iso-rotational excitation, whereas $a_9$ stems from the spin excitation. We remind the reader that this expression is derived under the assumption that the isospin excitation takes its minimum possible value, i.e. $i=|i_3|$, therefore it should be compared with the nucleus of lowest energy (highest binding energy) compatible with the quantum numbers $A$, $Z$ and $j$. 

\begin{figure}[h]
\begin{center}
\includegraphics[width=0.6\textwidth]{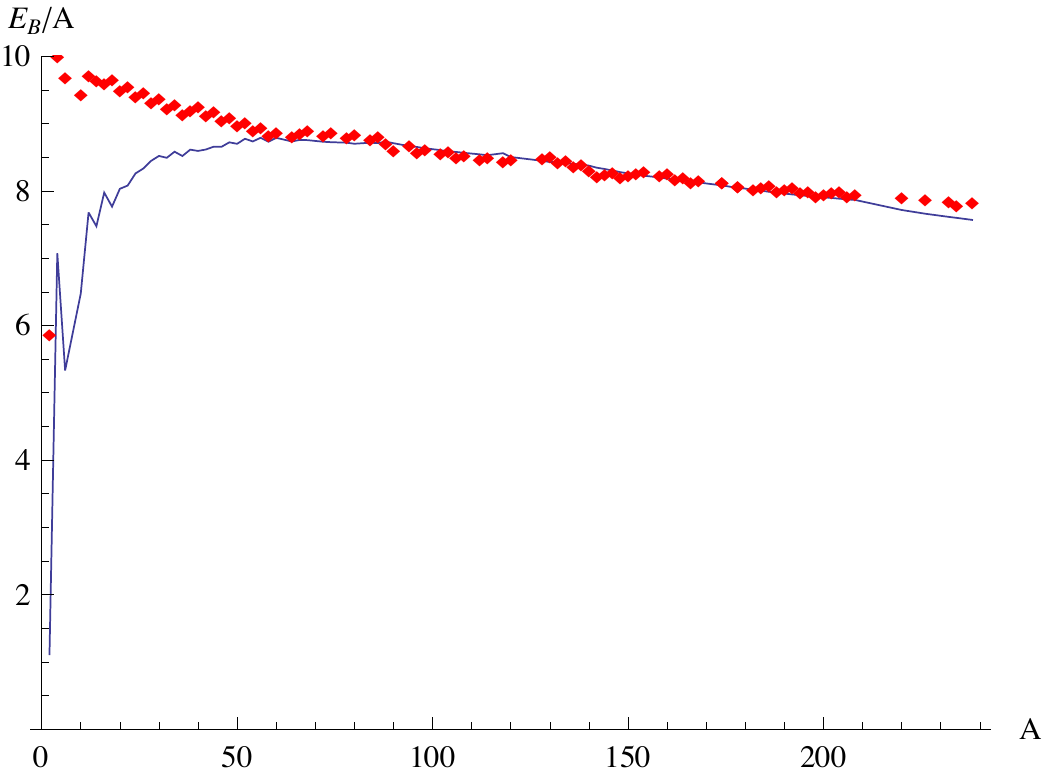}
\caption{Binding energies per nucleon in MeV. The experimental values are described by the solid line whereas the diamonds represent the results from our model.}
\label{FigEB}
\end{center}
\end{figure}

For a comparison with experimental values we now follow the strategy of \cite{marleau}. That is to say, for each fixed value of the atomic weight number $A$, we choose the values of $Z$ and $j$ corresponding to the most abundant nucleus. For the resulting nuclei we then compare the binding energy per atomic weight number, $E_{\rm B} / A$, with its experimental value. The result is shown in Fig. \ref{FigEB}. 

We find an excellent agreement for sufficiently large nuclei, whereas for small nuclei our model overestimates the binding energies. 

Finally, we want to compare our results with the ones from the semi-empirical mass formula (Weizs\"acker formula) \cite{Krane}
\be
 E_{\rm B}^{\rm W}(A,Z) = a_{\rm V} A - a_{\rm S} A^{2/3} - a_{\rm C} Z (Z-1) A^{-1/3} 
 - a_{\rm A} \frac{(A-2Z)^2}{A} + \delta (A,Z), 
\ee
where
\begin{eqnarray}
&& \nonumber \\ &&
\delta(n,Z) = \left\{ \begin{array}{cl}
a_{\rm P} A^{-3/4} & N \; \textrm{and} \; Z \; \textrm{even}, \\
0 & A \; \textrm{odd}, \\
- a_{\rm P} A^{-3/4} & N \; \textrm{and} \; Z \; \textrm{odd},
\end{array} \right.
\nonumber \\ && \nonumber \\ &&
a_{\rm V} = 15.5 \; \MeV, \quad a_{\rm S} = 16.8 \; \MeV, \quad a_{\rm C} = 0.72 \; \MeV,  \nonumber
\\ &&
a_{\rm A}= 23 \; \MeV, \quad a_{\rm P} = 34 \; \MeV \nonumber 
\end{eqnarray}
Here, the parameters $a_{\rm V}$, etc., are empirical constants, and the corresponding terms have the following meaning. $a_{\rm V} \;\ldots$ volume term; $a_{\rm S} \; \ldots $ surface term. These two terms are motivated by the liquid drop model (binding energy contributions for a classical liquid drop). $a_{\rm C} \; \ldots$ Coulomb energy. It is assumed that only the protons contribute to the Coulomb energy. $a_{\rm A} \; \ldots$ asymmetry term. This term is motivated by the Pauli principle, taking into account that the addition of further fermions of the same species requires them to have higher (kinetic) energy, because the lowest energy states are already occupied. $a_{\rm P} \; \ldots$ pairing term. When comparing the Weizs\"acker formula with our binding energy expression (\ref{bind-en}), we find two terms with a direct correspondence, namely the volume term $a_{\rm V} \sim a_1$, and the Coulomb term $a_{\rm C} \sim a_5$, whereas the remaining terms have a slightly different dependence on $A$ and $Z$.

\begin{figure}[h]
\begin{center}
\includegraphics[width=0.6\textwidth]{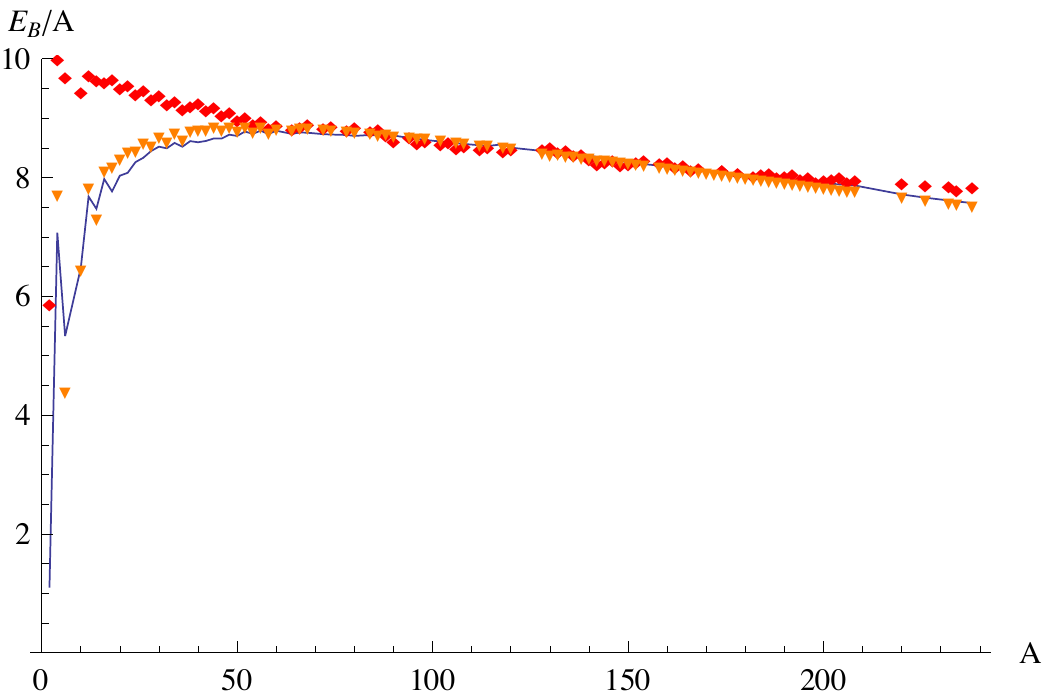}
\caption{Binding energy per nucleon in MeV from our model (diamonds) and Weizs\"acker's formula (triangles) compared to the experimental values (solid line).}
\label{FigWeizsacker}
\end{center}
\end{figure}

\begin{figure}[h]
\begin{center}
\includegraphics[width=0.6\textwidth]{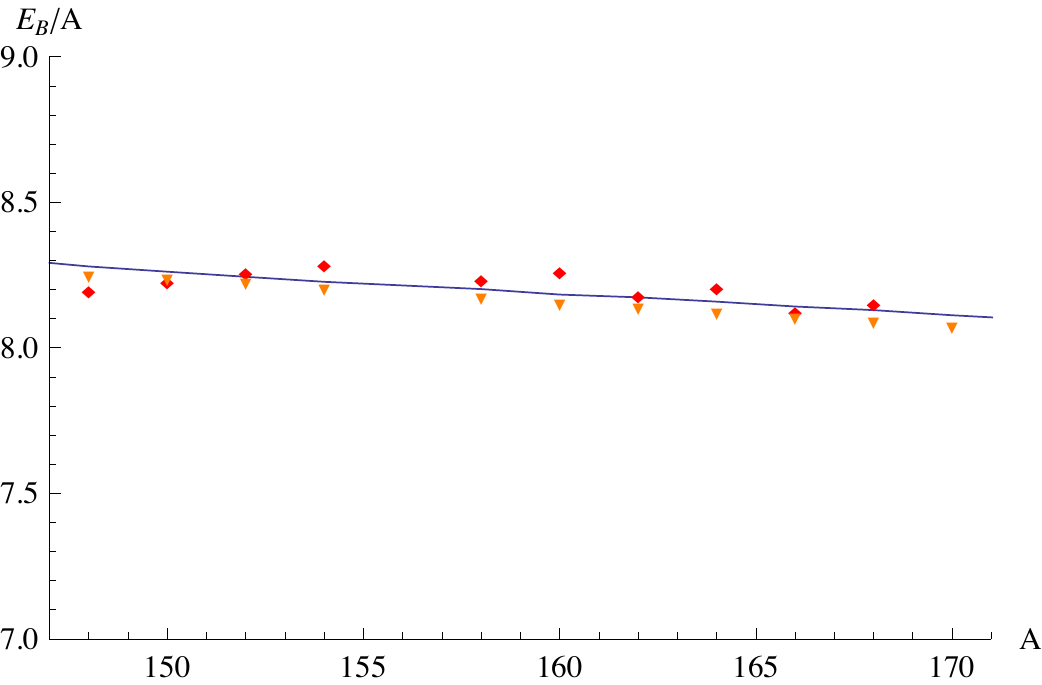}
\caption{Zoom of Fig. \ref{FigWeizsacker} for baryon numbers from 148 to 170.}
\label{FigZoom}
\end{center}
\end{figure}

The result of a comparison of both Weizs\"acker and our binding energies with the experimental values is shown in Fig. \ref{FigWeizsacker}. We may appreciate the following main differences between the results of the BPS Skyrme model and the Weizs\"acker formula. First of all, the Weizs\"acker formula describes very well also the binding energies of small nuclei. The main term responsible for this good behaviour is the surface term, which contributes to the binding energy per baryon number like $-a_S A^{-\frac{1}{3}}$ with the appreciable value $a_S \sim 17 \MeV$, significantly reducing the binding energies per $A$ for small values of $A$. Secondly, the BPS Skyrme model result shows some wiggles, i.e., rather sudden jumps for nearby nuclei even for large $A$, see Fig. \ref{FigZoom}. It is easy to understand the origin of these wiggles, which stem from the Coulomb energy contribution. 
Indeed, if we add a neutron to a given nucleus, then this has two effects on the Coulomb contribution to the binding energy. The Coulomb energy goes up, because the neutron has a (small) nonzero charge density, and the Coulomb energy goes down, because the radius of the nucleus increases. It turns out that the second effect is much stronger, so the Coulomb energy decreases (the binding energy increases) when a neutron is added. So, if we go from a nucleus with atomic weight number $A$ to another nucleus with $A+2$ by adding two neutrons, the binding energy goes up. On the other hand, if we go from $A$ to $A+2$ by adding a proton and a neutron, the binding energy goes down. Interestingly, this effect is not seen in the Weizs\"acker formula, although the contribution from the Coulomb term alone is even more pronounced (the Coulomb energy of the neutron is zero). The reason is that the Coulomb contribution to these jumps is balanced by the asymmetry term. Indeed, adding a neutron to a nucleus with neutron excess (i.e., any heavy nucleus) increases the asymmetry energy, compensating the decrease in Coulomb energy.

\section{Discussion}
In this article we studied the possibility to describe nuclear binding energies within our BPS Skyrme model. Let us emphasize, again, that there are strong theoretical arguments in favour of the assumption that this model is a good starting point for the description of some properties of nuclei. First of all, the classical Lagrangian of the BPS Skyrme model consists of two terms which both incorporate some collective properties of the underlying theory of strong interactions. The sextic term is just the square of the baryon current density with an obvious relation to the collective topological excitations of the underlying microscopic fields. The potential term, on the other hand, is supposed to describe the chiral symmetry breaking in the (pionic) effective field theory description. This point of view is further strenghtened by the BPS property of the BPS Skyrme model, which provides a natural explanation of the small binding energies of physical nuclei, and by the symmetries of the model, which are just the symmetries of an incompressible ideal liquid, explaining the viability of the nuclear liquid drop model. To keep calculations transparent, we just introduced the necessary minimum amount of structure in our investigations. Concretely, we started from the classical BPS soliton solutions of a certain (axially symmetric) ansatz, and then included the following additional features. i) The collective coordinate quantization of spin and isospin. This is absolutely necessary, because both spin and isospin are relevant quantum numbers of nuclei and nucleons and, therefore, indispensible for their description. ii) The Coulomb energies of the solitons. The Coulomb energy is expected to contribute significantly to the binding energies of higher nuclei, because of its fast growth with the nuclear charge $Z$. iii) An explicit breaking of the isospin symmetry which takes into account the mass difference between proton and neutron. 
The result of our investigation is that this natural, "minimal" version of the BPS Skyrme model already provides very good results for the description of nuclear binding energies for higher nuclei. Let us emphasize that the model contains just three fit parameters, and all of them are completely natural elements of the same underlying field theoretical description. The result that the model provides more accurate results for large $A$ is, in fact, to be expected, because a collective description in terms of a field theoretic version of the "nuclear liquid droplet" will be more adequate for higher nuclei. For small nuclei, single-particle properties or propagating pionic degrees of freedom should be more important, and their treatment will require an extension of the model, i.e., the inclusion of more terms into the lagrangian, and the study of additional effects.  

We remind the reader that in the course of our calculations we had to make certain choices at the beginning, in that we chose the standard Skyrme potential and fixed the symmetries of our soliton solutions to the axially symmetric ansatz (\ref{ansatz}). These choices do not influence the classical soliton energies (these are fitted to nucelar masses in any case), but they obviously do influence the spin ands isospin excitations (the moments of inertia) and the Coulomb energies. Here our point of view is that for the rather generic investigations on the viability of the BPS Skyrme model for the description of nuclear binding energies presented in this paper, these differences are not too important. That is to say, even if physical nuclei do not correspond to the axially symmetric field configurations assumed here, the resulting excitation and Coulomb energies will not be too different as long as the deviation from the axial symmetry (i.e., from a spherically symmetric baryonic density) is not too pronounced.  
For more detailed investigations, e.g., on nuclear spectroscopy (i.e., on possible excitated states of a given nucleus), on the other hand, the knowledge of the true symmetries of the corresponding skyrmion {\em is} important. We briefly comment on this issue below.

Before continuing our general discussion, we want to explain how our results relate to the ones of \cite{marleau}, where similar detailed computations have first been performed. Indeed, \cite{marleau}, too, start from the BPS Skyrme model, and their calculations are in many respects similar to ours. E.g., the multipole treatment of the Coulomb energy or the explicit isospin breaking are equivalent. There exist, however, several important differences. First of all, their expression for the isospin current density operator is completely different from our expression (\ref{iso-dens-op}), in that it depends only on the body-fixed isospin angular momenta and not on the collective coordinates which, we believe, is not correct. As a consequence, their results for the Coulomb energies are different. A second difference is that the authors of \cite{marleau} also include contributions from the terms ${\cal L}_2$ (sigma model term) and ${\cal L}_4$ (Skyrme term) to the classical soliton energies. While we completely agree that an improved treatment requires the inclusion of these terms, we do not agree with the specific way this is done in \cite{marleau}. The authors include these contributions in a perturbative fashion, by inserting the soliton solutions of the ${\cal L}_{06}$ model into the additional terms ${\cal L}_2$ and ${\cal L}_4$. A first subtlety of this procedure is related to the choice of potential. The standard (pion mass) quadratic Skyrme potential (\ref{Sk-pot}) leads to compact soliton solutions, and these solutions produce singular expressions when inserted into the energy functional corresponding to ${\cal L}_2$. The standard Skyrme potential is, therefore, not adequate for this perturbative calculation. The authors of \cite{marleau} are, of course, fully aware of this problem and choose different potentials which lead to non-compact solitons. Now their procedure simply consists in inserting the soliton solutions of the restricted model ${\cal L}_{06}$ resulting from the axially symmetric ansatz (\ref{ansatz}) into the energy functionals corresponding to ${\cal L}_2$ and ${\cal L}_4$. The problem is that the resulting perturbative energy contributions grow quite fast with the atomic weight number $n=B=A$. Specifically, a contribution from ${\cal L}_2$ grows like $A^\frac{7}{3}$. For the authors of \cite{marleau}, this is, in fact, a wellcome result, for the following reason. As is obvious, e.g., from Fig. \ref{FigEB}, the binding energy per atomic weight number $A$ should go down for large values of $A$. The physical expectation is that this decrease is largely due to an increase of the Coulomb energy per atomic weight number for large nuclei, and this is precisely what we find in our calculations. The authors of \cite{marleau} with their different results for the Coulomb energies, however, find that only about one-half of this decrease comes from the Coulomb energy, whereas the other half comes from the increase of the perturbative energy contributions per atomic weight number $A$ for large $A$. This fast growth of the perturbative energies with the baryon number is, however, not acceptable from a theoretical point of view. Indeed, although perturbative, these contributions are part of the classical soliton energies, and soliton energies can never grow faster than linear in the topological degree, because otherwise the "solitons" would be completely unstable already at the classical level. This does not imply that the perturbative treatment per se is not viable, it just means that the axially symmetric ansatz (\ref{ansatz}) is not the right choice for a perturbative calculation. We remind the reader that due to the large (in fact, infinite-dimensional) symmetry group of the BPS Skyrme model, there exist infinitely many more BPS solutions with the same energy with rather arbitrary shapes. The proper procedure then should consist in minimizing the perturbative energy contributions over all these infinitely many BPS solutions and in choosing the corresponding minimizer in each topological sector \cite{speight2}.  Thirdly, we restricted our discussion to nuclei with even baryon number, because, in our opinion, the Finkelstein-Rubinstein constraints for skyrmions with axial symmetry require this restriction, whereas there is no such restriction in \cite{marleau}.   

Altoghether, we have found that already a rather "minimal" version of the BPS Skyrme model allows for a very accurate description of nuclear binding energies. Still, these results just constitute some first steps of what might quite probably become a rather extended field of investigation.  
One first generalization simply consists in considering different potentials and in studying their influcence on the physical properties of the resulting "nuclei". As emphasized already, without the sigma-model term, there is no direct relation of the standard Skyrme potential to pion masses and, therefore, no obvious reason to prefer it over other potentials. In addition, as mentioned in the previous paragraph, there exists the problem that with the standard Skyrme potential a perturbative inclusion of the sigma model term is not possible. This seems to constitute a kind of contradiction, because with the sigma-model term present, the standard Skyrme potential does have the pion mass interpretation and its presence is required on physical grounds. A possible resolution is provided by the following proposal. One starts with a BPS submodel
\be
{\cal L}_{\rm BPS}={\cal L}_6 + \tilde {\cal L}_0
\ee
 where the potential $\tilde{\cal L}_0$ is {\em different} from the standard Skyrme potential and should be motivated by some physical or phenomenological requirements (e.g., reproducing the rather flat baryon density profiles of physical nuclei). Then one includes both the sigma-model term ${\cal L}_2$ and the standard Skyrme potential (\ref{Sk-pot}) at the same level (e.g., perturbatively,)
\be
{\cal L} = {\cal L}_{\rm BPS} + \epsilon ({\cal L}_2 +  {\cal L}_0 ).
\ee
This has the additional advantage that the relative strengths of sigma model and Skyrme potential terms may be fixed to their physical value (i.e., reproducing the physical pion mass) without interfering with the BPS property of the remaining terms.  The results of the present paper correspond in a certain sense to the limit $\epsilon \to 0$, however, with the choice $\tilde {\cal L}_0 = {\cal L}_0$. Going beyond this limit in a perturbative fashion requires,  
as already briefly mentioned, the determination of the minimizers of the sigma model term among all possible solutions (i.e., among all VPD orbits of a given solution) of the BPS submodel in each topological sector, \cite{speight2}. Due to the infinitely many symmetries of the BPS Skyrme model and the complicated geometry of the VPDs, this minimization constitutes a rather nontrivial and interesting variational problem on its own, which will require a dedicated research program and the expertise of mathematicians. The semi-classical quantization of spin and isospin and the inclusion of both the Coulomb energy and the explicit isospin breaking should then be done equivalently to the present paper, but using the new classical solutions (the minimizers of the sigma model term). Physically, the minimization should improve the classical soliton energies and, even more importantly, it will allow to determine the correct symmetries of the corresponding solitons. These symmetries are so important because they, in turn, determine the Finkelstein-Rubinstein constraints which the corresponding nuclear wave functions have to obey. These Finkelstein-Rubinstein constraints determine the allowed and forbidden spin and isospin excitations and are, therefore, of utmost importance for the application of skyrmions to nuclear spectroscopy \cite{wood}. A further research direction is related to the possibility to quantize additional degrees of freedom in the BPS Skyrme model. Some first steps in this direction have been performed in \cite{vibr}, in the context of low energy hadron physics rather than nuclear physics, but a much more detailed investigation is certainly required. 

We conclude that we found strong evidence for the claim that our resricted BPS Skyrme model is a most
adequate starting point to fully implement the original program of Skyrme of a
unified field theoretic description, including geometrical and
topolgical elements, of low energy strong interaction physics. 
We believe it a worthwile enterprise to reanalyse physical systems where Skyrme theory has already been used successfully, departing from this new starting point, along the lines indicated in the previous paragraph. Such a new analysis might both complement existing calculations and lead to a significant quantitative improvement of existing results in nuclear and hadron physics. 
Besides the gratifying conceptual aspects of the 
proposal, which explain in a simple way the symmetries and dynamics,
it offers promising progress due to the analytical and exact results
it enables, both in basic understanding  and useful phenomenology.



\section*{Appendix A}

The auxiliary matrices for the Weyl-ordered isospin current density operator (\ref{iso-dens-op}) are
\begin{equation} \label{Aip}
{\cal A}_{ij}=\left( \begin{array}{ccc}
n^2 \cos^2(n \phi)+\cos^2 \theta \sin^2(n\phi) & \frac{1}{4}(1-2n^2+\cos(2\theta)) \sin(2n\phi) & \sin \theta \cos \theta \sin(n\phi) \\
&& \\
\frac{1}{4}(1-2n^2+\cos(2\theta)) \sin(2n\phi) & \cos^2 \theta \cos^2(n\phi)+n^2 \sin^2(n\phi) & \sin \theta \cos \theta \cos(n\phi) \\
&& \\
 \cos \theta \sin \theta \sin(n\phi) & \sin \theta \cos \theta \cos(n\phi) & \sin^2 \theta \end{array} \right).
\end{equation}

\begin{eqnarray} \label{Bir}
&{\cal B}_{ij} = & n \left( \begin{array}{ccc}
- n \cos \phi \cos(n\phi) - \cos^2 \theta \sin \phi \sin(n\phi) & n \cos(n\phi)\sin \phi - \cos^2 \theta \cos \phi \sin(n\phi) & \\
&& \\
 -\cos^2 \theta \sin \phi \cos(n\phi)+n\cos \phi \sin(n\phi) & \; - \cos^2 \theta \cos \phi \cos(n\phi) - n \sin \phi \sin(n\phi) & \cdots \\
&& \\
- \sin \theta \cos \theta \sin \phi & -  \sin \theta \cos \theta \cos \phi & \end{array} \right. \nonumber \\
&& \nonumber \\ 
&& \nonumber \\ 
&& \left. \begin{array}{cc}
& - \sin \theta \cos \theta \sin(n\phi) \\
& \\
\cdots & - \sin \theta \cos \theta \cos(n\phi) \\
& \\
& - \sin^2 \theta \end{array} \right).
\end{eqnarray}

\section*{Appendix B}

Here we demonstrate that the non-diagonal matrix elements of the electric charge density vanish. We prefer to work with Euler angles, where
\begin{equation}
K_1 = - \im \left( \frac{\cos \gamma}{\sin \beta} \frac{\partial}{\partial \alpha} - \sin \gamma \frac{\partial}{\partial \beta} - \cot \beta \cos \gamma \frac{\partial}{\partial \gamma} \right), \nonumber
\end{equation}
\begin{equation}
K_2 = - \im \left( - \frac{\sin \gamma}{\sin \beta} \frac{\partial}{\partial \alpha} - \cos \gamma \frac{\partial}{\partial \beta} + \cot \beta \sin \gamma \frac{\partial}{\partial \gamma} \right),
\end{equation} 
\begin{equation}
K_3 =  \im \frac{\partial}{\partial \gamma}, \nonumber
\end{equation}
for the body-fixed isospin operator. Further,
\begin{equation}
R_1 = - \cos \gamma \sin \beta, \qquad \qquad R_2 = \sin \gamma \sin \beta, \qquad \qquad R_3 = \cos \beta
\end{equation}
and, as a consequence,
\be
[K_i , R_j ]= \im \epsilon_{ijk}R_k .
\ee
Both the spin and the isospin factors of the nuclear wave functions are related to the (complex conjugates of the) Wigner D matrices. We are interested only in the isospin part,  where Wigner's D matrices are given by
\begin{equation}
D^{(i)}_{i_3 k_3} (\alpha, \beta, \gamma) = e^{-\im i_3 \alpha} d^{(i)}_{i_3 k_3}(\beta) e^{-\im k_3 \gamma}
\end{equation}
(see \cite{Dmatrix} for details, and for the definition of the $d^{(i)}_{i_3 k_3}$). 
If we make the following choice for the (isospin) nuclear wave functions,
\be
|ii_3k_3 \rangle = (-1)^{k_3} D^{(i)*}_{i_3,-k_3} ,
\ee
then the action of $\vec K$ on the wave functions is exactly equivalent to the action for the standard representation of angular momentum,
\be
K_3 |ii_3k_3 \rangle = k_3 |ii_3 k_3\rangle \; , \quad (K_1 \pm iK_2 ) |ii_3k_3 \rangle = \sqrt{i(i+1) - k_3(k_3 \pm 1)}|ii_3k_3\pm 1 \rangle
\ee
(remember that $\vec K$ in our conventions obeys the standard angular momentum algebra and is related to the space-fixed isospin by {\em minus} a rotation). 
Further, in the cases of interest we always have $k_3=0$, where the D matrix simplifies to
\begin{equation}
D^{(i)}_{i_3 0} (\alpha, \beta, \gamma) = \sqrt{\frac{4 \pi}{2i + 1}} Y^*_{ii_3} (\beta, \alpha)
\end{equation}
and $|ii_3 0\rangle =D^{(i)*}_{i_3 0}$. For reasons of symmetry, it is sufficient to check the following three matrix elements (we suppress $k_3=0$) $\langle ii_3 |K_1 R_2 + R_2 K_1|ii_3 \rangle $, $\langle ii_3 |K_1 R_3 + R_3 K_1 |ii_3 \rangle $ and $\langle ii_3 |K_3 R_1 + R_1 K_3 | ii_3 \rangle $.
Here, the first case is the most difficult one,
\be
\langle ii_3 |K_1 R_2 + R_2 K_1|ii_3 \rangle = \langle ii_3 |2R_2 K_1 + [K_1 ,R_2] |ii_3 \rangle .
\ee
Using $[K_1 ,R_2] =\im R_3 = \im \sqrt{4\pi /3} Y_{10}$, we find for the second, commutator term
\be
\langle ii_3 | [K_1 ,R_2] |ii_3 \rangle = \frac{(4\pi)^\frac{3}{2} \im }{\sqrt{3}(2i+1)}\int_0^{2\pi} d\alpha \int_0^{2\pi}d\gamma \int_0^\pi \sin \beta d\beta
Y^*_{ii_3}Y_{10}Y_{ii_3} =0. 
\ee
The operator $K_1$ in the first term contains derivatives w.r.t. $\alpha$, $\beta$ and $\gamma$. Here the $\gamma $ derivative does not contribute because $Y_{ii_3}(\beta ,\alpha)$ does not depend on it. The $\alpha$ derivative does not contribute  because it is multiplied by $\sin \gamma \cos \gamma$ which integrates to zero. The remaining $\beta$ derivative leads to
\be
\langle ii_3 |2R_2 K_1|ii_3 \rangle = \frac{-8\im \pi}{2i+1} 
\int_0^{2\pi} d\alpha \int_0^{2\pi}d\gamma \int_0^\pi \sin \beta d\beta \sin^2 \gamma Y^*_{ii_3}\sin\beta \partial_\beta Y_{ii_3} =0.
\ee
The easiest way to see that this integral is zero, indeed, is by expressing the $Y_{ii_3}$ by the associated Legendre functions, $Y_{ii_3}(\beta ,\alpha ) = c_{ii_3} e^{\im i_3 \alpha}P_{ii_3} (t)$ where $t = \cos \beta$ and $\sin \beta \partial_\beta = -(1-t^2)\partial_t$, and by using the recurrence formula
\be
(1-t^2)\partial_t P_{lm}(t) = \frac{1}{2l+1}\left( (l+1)(l+m)P_{l-1,m}(t) + l(l-m+1)P_{l+1,m}(t)\right) .
\ee
The result then follows from the orthogonality relations of the associated Legendre functions.

In the remaining matrix elements $\langle ii_3 |K_1 R_3 + R_3 K_1 |ii_3 \rangle $, etc.,  always one operator index is equal to 3, whereas the other one takes the values 1 or 2. As a consequence, there always appears precisely one factor of $\sin \gamma$ or $\cos \gamma$ in the resulting integrand, which integrates to zero. 

\section*{Acknowledgement}
The authors acknowledge financial support from the Ministry of Education, Culture and Sports, Spain (grant FPA2008-01177), 
the Xunta de Galicia (grant INCITE09.296.035PR and
Conselleria de Educacion), the
Spanish Consolider-Ingenio 2010 Programme CPAN (CSD2007-00042), and FEDER. 
CN thanks the Spanish
Ministery of
Education, Culture and Sports for financial support (grant FPU AP2010-5772).
Further, AW was supported by polish NCN grant 2011/01/B/ST2/00464. The authors thank M. Speight, N. Manton and R. Vazquez for helpful discussions.


\begin{thebibliography}{99}

\bibitem{skyrme} T. H. R. Skyrme, Proc. Roy. Soc. Lon. {\bf 260},
127 (1961); Nucl. Phys. {\bf 31}, 556 (1962); J. Math. Phys. {\bf
12}, 1735 (1971).
\bibitem{manton}
N. S. Manton, 
Commun. Math. Phys. {\bf 111}, 469 (1987); 
C. J. Houghton, N. S. Manton, P. M. Sutcliffe,
Nucl. Phys. B{\bf 510}, 507 (1998);
R. A. Battye, P. M. Sutcliffe, Nucl. Phys. B{\bf 705}, 384 
(2005); R. A. Battye, P. M. Sutcliffe, Phys. Rev. C{\bf 73},  055205 (2006); 
R. A. Battye, N. S. Manton, P. M. Sutcliffe,
Proc. Roy. Soc. Lond. A{\bf 463}, 261 (2007);
D.T.J. Feist, P.H.C. Lau, N.S. Manton, Phys. Rev. D{\bf 87}, 085034 (2013).
\bibitem{thooft} G. t'Hooft, Nucl. Phys. B{\bf 72}. 461 (1974); E. Witten, Nucl. Phys. B{\bf 160}, 57 (1979); E. Witten, Nucl. Phys. B{\bf 223}, 433 (1983).
\bibitem{nappi} G. S. Adkins, C. R. Nappi, E. Witten, Nucl. Phys. 
B{\bf 228}, 552 (1983); G. S. Adkins, C. R. Nappi, Nucl. Phys. B{\bf 233}, 109 
(1984).
\bibitem{braaten}
E. Braaten, L. Carson,
Phys. Rev. Lett. {\bf 56}, 1897 (1986); 
Phys. Rev. D{\bf 38}, 3525 (1988).
\bibitem{carson}
L. Carson,
Phys. Rev. Lett. {\bf 66}, 1406 (1991); 
L. Carson,
Nucl. Phys. A{\bf 535}, 479 (1991);   
T.S. Walhout,
Nucl. Phys. A{\bf 531}, 596 (1991). 
\bibitem{wood}
O. V. Manko, N. S. Manton, S. W. Wood,
Phys. Rev. C{\bf 76}, 055203 (2007); 
R. A. Battye, N. S. Manton, P. M. Sutcliffe, 
S. W. Wood, Phys. Rev. C{\bf 80}, 034323 (2009).
\bibitem{rybakov-book}
V.G. Makhankov, Y.P. Rybakov, V.I. Sanyuk, "The Skyrme Model", Springer Verlag, Berlin, 1993.
\bibitem{manton-book}
N. Manton, P. Sutcliffe, "Topological Solitons", Cambridge University Press, Cambridge, 2007.
\bibitem{BPS-Sk}
C. Adam, J. Sanchez-Guillen, A. Wereszczynski,
Phys. Lett. B{\bf 691}, 105 (2010)
[arXiv:1001.4544];
C. Adam, J. Sanchez-Guillen, A. Wereszczynski,
Phys. Rev. D{\bf 82}, 085015 (2010)  
[arXiv:1007.1567].
\bibitem{marleau1}
E. Bonenfant, L. Marleau,
Phys. Rev. D{\bf 82}, 054023 (2010).
\bibitem{fosco}
C. Adam, C. D. Fosco, J. M. Queiruga, J. Sanchez-Guillen, A. Wereszczynski,
J. Phys. A{\bf 46}, 135401 (2013). 
\bibitem{sutcliffe}
P. Sutcliffe, JHEP {\bf 1008}, 019 (2010); JHEP {\bf 1104}, 045 (2011).
\bibitem{rho} Y.-L. Ma, Y. Oh, G.-S. Yang, M. Harada, H. K. Lee,
B.-Y. Park, M. Rho, Phys. Rev. D{\bf 86}, 074025 (2012); Y.-L. Ma, G.-S.
Yang, Y. Oh, M. Harad, Phys. Rev. D{\bf 87}, 034023 (2013).
\bibitem{ChPT}
J. Gasser and H. Leutwyler, Ann. Phys. (N.Y.) {\bf 158} (1984) 142; I. J.
R. Aitchison, C. M. Fraser, Phys. Rev. D{\bf 31} (1985) 2605; I. J. R.
Aitchison, C. M. Fraser,
P. J. Miron, Phys. Rev. D{\bf 33} (1986) 1994; J. Zuk, Z. Phys. C{\bf 29} (1985)
303; C. M. Fraser, Z. Phys. C{\bf 28} (1985) 101; I. J. R. Aitchison, C. M.
Fraser, E. Tudor, J. Zuk, Phys. Lett. B{\bf 165} (1985) 162.
\bibitem{sextic} A. Jackson, A. D. Jackson, A. S. Goldhaber,
G. E. Brown, L. C. Castillejo, Phys. Lett. B{\bf 154}, 101 (1985);
L. Floratos, B. M. A. G. Piette, Phys. Rev. {\bf D64}, 045009 (2001);  V. Kopeliovich, Phys. Part. Nucl. {\bf 37}, 623 (2006).
\bibitem{ferr}
C. Adam, L. A. Ferreira, E. da Hora, A. Wereszczynski, W. J. Zakrzewski,
e-Print: arXiv:1305.7239 [hep-th].
\bibitem{speight}
J. M. Speight, 
J. Phys. A{\bf 43}, 405201 (2010). 
\bibitem{potentials} V. B. Kopeliovich, B. Piette, W. J. Zakrzewski,
Phys. Rev. D{\bf 73}, 014006 (2006); B. Piette, W. J. Zakrzewski, Phys. Rev.
D{\bf 77}, 074009 (2008); C. Adam, C. Naya, J. Sanchez-Guillen, A.
Wereszczynski, Phys. Rev. D{\bf 86}, 085001 (2012). 
\bibitem{marleau}
E. Bonenfant, L. Harbour, L. Marleau,
Phys. Rev. D{\bf 85}, 114045 (2012); M.-O. Beaudoin, L. Marleau, arXiv:1305.4944. 
\bibitem{Fin-Rub}
D. Finkelstein, J. Rubinstein,
 J. Math. Phys. {\bf 9}, 1762 (1968).
\bibitem{krusch}
S. Krusch, Ann. Phys. {\bf 304}, 103 (2003) [hep-th/0210310];
Proc. Roy. Soc. Lond. A{\bf 462}, 2001 (2006) [hep-th/0509094].
\bibitem{hou-mag}
C. Houghton, S. Magee, Phys. Lett. B{\bf 632}, 593 (2006).
\bibitem{cal-wit}
C.G. Callan, E. Witten, Nucl. Phys. B{\bf 239}, 161 (1984).
\bibitem{Ding}
G.-J. Ding, M.-L. Yan, 
Phys. Rev. C{\bf 75}, 034004 (2007).
\bibitem{iso-breaking}
E.~Rathske,
Z. Phys. A\textbf{331}, 499 (1988); 
P.~Jain, R.~Johnson, N.~W.~Park, J.~Schechter and
H.~Weigel,
Phys. Rev. D\textbf{40}, 855 (1989);
U.~-G.~Meissner, A.~M.~Rakhimov, A.~Wirzba and
U.~T.~Yakhshiev,
EPJ Web Conf.\ \textbf{3}, 06008 (2010) [arXiv:0912.5170].
\bibitem{Masses} G. Audi, A. H. Wapstra and C. Thibault, Nucl. Phys. A{\bf 729}, 337 (2003).

\bibitem{Krane} K. S. Krane, {\it Introductory Nuclear Physics}, John Wiley \& Sons, 1988.
\bibitem{speight2}
J. M. Speight, private communication.
\bibitem{vibr}
C. Adam, C. Naya, J. Sanchez-Guillen, A. Wereszczynski,
arXiv:1306.6337 [hep-th].
\bibitem{Dmatrix} A. deShalit and H. Feshbach, {\it Theoretical Nuclear Physics Volume I: Nuclear Structure}, John Wiley \& Sons, 1974.

\end{thebibliography}
\end{document}